\documentstyle[]{l-school}
\def\lb{\hfil\break}
\def\to{\rightarrow}
\def\del{\partial}
\def\Phivec{\vec\Phi}
\def\half{{1\over 2}}
\def\tr{\mathop{\rm tr}\nolimits}
\begin{document}
\markboth{M. Moshe}
{Quantum F.T. in Singular Limits}
\setcounter{part}{16}

\title{ QUANTUM FIELD
THEORY IN  SINGULAR LIMITS
\footnote{\it  A Short Summary - Les-Houches February 1997}
}
\author{Moshe Moshe
\footnote{\it  This  research
was supported in part by a grant from the Israel
Science Foundation}
}
\institute{Department of Physics\\
Technion - Israel Inst. of Technology\\
Haifa, 32000  ISRAEL}
\maketitle

\section{INTRODUCTION}

It has been suggested \cite{Kazakov} to approximate
geometry by dense Feynman graphs of
the same topology, taking
the number of vertices to infinity.
The dynamically triangulated random surfaces summed
on different topologies are then viewed as the manifold for
string propagation. Summing up
Feynman graphs of the O(N) matrix model in d dimensions
 results in a genus expansion and it provides, in some sense,
a nonperturbative treatment of  string theory
when the  double scaling limit is enforced \cite{Gross}.
If these efforts could be extended to d $>1$ dimensions,
then a major progress would have been  achieved in
studying a long lasting problem
in elementary
particle theory. Namely, the
relation between d dimensional quantum field theory
and its possible formulation in terms of strings.
The possibility of reaching a "stringy" representation of
SU(N) and U(N) quantum field theory in a
correlated singular limit
was proposed some time ago in \cite{Neuberger}.

$O(N)$ symmetric vector models represent discretized
filamentary surfaces
- randomly branched polymers
in the double scaling limit,
in the same manner in which matrix models, in
their double scaling limit,
 provide representations  of
dynamically triangulated random
surfaces.
Though  matrix theories attract
most  attention, a detailed understanding of these theories exists
only for dimensions  $d \leq 1$. On the other hand,
in  many cases, the double scaling limit of the $O(N)$ vector
 models\cite{Nishigaki,Zinn} can be successfully studied  also  in
dimensions $d > 1$
thus, providing
intuition for the search
for a possible description of quantum field theory
in terms of extended objects,
string-like excitations, in four dimensions.
This  stimulated  new studies \cite{Zinn}-\cite{Eyal} of the phase
structure of O(N) vector quantum field theories that offered
a new look into this long lasting problem.
Most often,  quantum field  theories
are described by  their  point-like  quanta
 and  a direct relation to a string-like
structure is difficult to visualize; if there is a correspondence
to an hadronic string, it must be of a less
direct kind \cite{Neuberger}.

\section{SINGULAR LIMITS OF  O(N) SYMMETRIC MODELS}
\subsection{The Double Scaling Limit in O(N) Matrix Models}

In O(N) matrix models the double scaling limit is enforced
in the calculation of the partition function
$$ Z_N(g)~=~\int D\hat\Phi exp\{-\beta\int d^dx Tr\{\hat\Phi(x)
\hat K \hat\Phi(x)
+V(\hat\Phi(x))\} \}   \eqno [1]$$
\noindent where $\hat\Phi(x)$ is an NxN
Hermitian matrix and V($\hat\Phi(x)$) is the potential
depending on the coupling(s)
constant(s) $\{ g_i\}$.

In zero dimensions,
after performing the integration on the angular variables,
one is left with the  integration on the eigenvalues $\lambda_i$
$$ Z_N(g)~=~\Omega_N\int\prod_{i}^{N}d\lambda_i
~~exp \bigl\{2\sum_{i,j}^{N} \ln |\lambda_i- \lambda_j | - \beta
\sum_{i}U(\lambda_i)\bigr\}. \eqno [2]$$
In Eq. [2] one notes a Pauli repulsion between the eigenvalues, and  a
critical point $\{ g_i \} = \{ g_{iC}\} $ is
found when the Fermi level reaches the extremum of
the potential. The weak coupling limit,namely,
${1\over N} \sim{1\over \beta}\to 0$, gives a one-dimensional
frozen Dyson gas and the planar graphs dominate.
As $ N\to \infty $ and ${N\over\beta} \to 1$  (or
$\{ g_i \} \to  \{ g_{iC}\}$),
the "melting" of the gas starts and the
non-planar graphs become important.
The genus (G) expansion  of the free energy of the
system is given by
($S$ is the area which is proportional to the order
of the Feynman graphs):
$$F =\ln Z_N~=~a~+~b\ln\beta~+~\sum_{G,S}
N^{2(1-G)}~~({N\over\beta})^S ~~F_S
~~\sim \sum_G ({1 \over N})^{2G-2}
{\cal A}_G \{g_i\}
\eqno [3]$$

This topological series  is  not Borel summable
 due to the factorial growth of the positive
${\cal A}_G \{g_i\}$ with the genus G,
and a nonperturbative approach is needed.
As the set of coupling constants $\{g_i\}
\longrightarrow \{g_{iC}\}$ approaches a set of critical
values, the loop expansion at a given topology diverges.
At a given topology, ${\cal A}_G \{g\}$
has a finite radius of convergence when expanded
in powers of the coupling constant g.
The limit  $g \rightarrow g_c$
emphasizes the higher order terms in this
expansion, namely, the
denser Feynman graphs.
Typically, for a potential $V(\hat\Phi)$ with
a single coupling,
$${\cal A}_G \{g\} \sim (g-g_C)^{-\kappa_G} \longrightarrow
\infty ~~~\{ {\rm ~as~} g \rightarrow
g_c \} \eqno [4]$$

In the suitable singular limit
 (the double scaling limit),
all terms in the topological expansion in
Eq. [3] are of equal importance.
The nonperturbative framework needed here should be capable
of reproducing the topological series as an asymptotic expansion
but should also be a framework for
non-perturbative calculations\cite{Gross}.
In the limit $\beta$=N
$\rightarrow \infty$ and $g \rightarrow
g_c$ in a correlated manner, the powers of
${1\over N}$ in  Eq. [3] are compensated by the
growing ${\cal A}_G \{g\}$ in Eq.[4]. In this limit, Eq.[3]
turns into
an expansion in the string coupling  constant
since all genera are relevant now. The physical
meaning of this formal limit will be soon  clarified in
the O(N) vector quantum field theory.

The model belongs to the same universality class of two
dimensional conformally invariant matter field in gravitational
background and thus, the critical exponents of this model
are calculable from Eq. [3]. One may, indeed, find it quite
remarkable that these relatively simple
integrals possess the critical behavior
of two dimensional gravity coupled to conformal matter.

\subsection{The Double Scaling Limit in O(N) Vector Models}

The double scaling limit of the O(N) vector models
follows an analogous
procedure,
$$ Z_N(g)~=~\int {\cal D}\vec\Phi
exp\{-\beta\int d^dx \{\vec\Phi(x)
\hat K \vec\Phi(x)
+V(\vec\Phi^2(x))\} \}\eqno[5]$$
where $\vec\Phi$ is an O(N) vector. In zero dimensions,
this turns into
a simple N dimensional integral over the components
of $\vec\Phi$.
The expansion in terms of Feynman graphs in the
large N limit resembles now  an expansion  of
"randomly branched
polymers"\cite{Nishigaki}. The graphical realization
is obtained when
the dual diagrams are defined from the Feynman
graphs by interchanging
the vertex and propagator in the Feynman graph
by the bond and "molecule"
in the dual graph, respectively, which describes
now the branched polymer.
$$\ln Z_N~=~\sum_{h,b} N^{(1-h)}~~
({N\over\beta})^b ~~F_b
~~~~~\sim \sum_h ({1 \over N})^{h-1}
{\cal A}_h \{g_i\}
\eqno[6]$$\smallskip
Following  here the suitable scaling procedure,
 Eq. [5] turns into
an expansion in  a "polymer" coupling  constant.
For $V(\vec\Phi^2) \sim g\vec\Phi^2(x)^2$
in the correlated limit $N\to\infty$ and $g\to g_C$, the
negative powers of N are compensated by
${\cal A}_h \{g\} \sim (g-g_C)^{-\gamma_h}
\longrightarrow
\infty ~~~\{ {\rm ~as~} g \rightarrow
g_c \}$.
Criticality comes  from the tuning
of the potential in order to balance the
centrifugal barrier. Whereas matrix models in
dimensions  $d > 1$
are very difficult to  study, even in the
leading large N limit
or the double scaling
limit, vector models  are relatively easy to
analyze in these limits even at
$d\geq 2$.

There are several interesting physical questions
that can be answered
when the formal double scaling limit is enforced
in $d\geq 2$ O(N) vector theories.
A possible list of questions follows: \lb
{\bf (1)} What is the
 phase structure, spectrum and symmetries of the quantum
field theory in this singular limit ?\lb
{\bf (2)} Does the flow
$g \to g_C$ agree with the renormalization group flow ? \lb
{\bf (3)} How are the ultraviolet divergences
taken into account at d$>$2 ~? (Needless to say that
these questions
were not raised in  matrix models in d$\leq$1). \lb
{\bf (4)} Taking into account that
we are discussing a renormalizable quantum field theory,
what is the nature of
 the divergences  needed for compensating
the powers of $N^{-h}$ in Eq. [6] so that all orders in
$N^{-1}$ are of equal importance?

These issues have been
discussed in \cite{Zinn}-\cite{Eyal}, while considering
as an example the large N limit of the self
interacting scalar O(N)
symmetric vector model in $0\leq{\rm d}<4$  Euclidian
dimensions
defined by the functional integral in Eq.[5] with
$\hat K = \del^2 D({-\del^2\over \Lambda^2})  $,
(where D(z) is a positive non-vanishing polynomial with D(0)=1).
The following results have been found \cite{DiVecchia,Eyal}:
\lb
{\bf (1)} The physical meaning of the double scaling
limit can be phrased as
tuning the force between the O(N) quanta so that
a singlet massless bound state is created in the
spectrum. This is
a singular limit in the sense that while $N \to \infty$
the coupling constant is often tuned to a negative value
$g \to g_C$. The physical spectrum consists of
the propagating O(N)  quanta of small
mass m in addition to the massless O(N) singlet. \lb
{\bf (2)} The flow of $g \to g_C$ is consistent with
the renormalization group flow, provided the limits
$N\to \infty$ and the cutoff $\Lambda \to \infty$ are
appropriately correlated. This point was made also
in Ref.\cite{Neuberger}.\lb
{\bf (3)} The ultraviolet divergences are dictating the
effective field theory obtained in the singular limit
of the double scaling limit. They enforce a detailed relation
between  $\Lambda$ and N, mentioned in (2) above,
 $\Lambda = \Lambda (N)$.\lb
{\bf (4)} The divergences that compensate the
decreasing powers of $N^{-h}$ in Eq. [6] and make all
orders in $N^{-1}$ of equal importance, are infrared
divergences.
Thus, in the double scaling limit, the tuning of
the forces that produce a massless singlet excitation,
produce infrared divergences which are the essential
ingredient of this limit.\lb
{\bf (5)}Following  (4) above, in order that the
compensating infrared singularities will show up,
the effective field theory of the singlet bound state is
super-renormalizable. \lb
{\bf (6)} In the critical dimensions (e.g. ,
$\Phi^6$ in d=3) the massless O(N) singlet excitation is
the Goldston boson
of spontaneous breaking of scale invariance - the
dilaton\cite{Bardeen}.

In the conventional treatment
of the large N limit,  Eq. [5] is expressed by:
$$ Z_N  =\int {\cal D}\rho\int {\cal D}\lambda
 ~~exp\bigl\{-N\int d^d x \,
[V(\rho)-\half \lambda
\rho]+\half N \tr\ln(-\del^2+\lambda ) . \eqno[7]$$
For $N$ large  the integral is evaluated by the steepest descent.
The saddle point value $\lambda$ is the
$\vec\Phi$-field mass squared
 $\lambda=m^2$ and $\rho =\rho_s$ at the saddle point.
The matrix of the  second partial derivatives
of the effective action
is:
$$N\pmatrix{ V''(\rho) & -\half \cr
-\half & -\half B_2(p;m^2)  \cr}  ,  \eqno[8]$$
where $B_2(p;m^2)$ is the appropriate "bubble graph".
Since the integration contour for $\lambda=m^2$
should be parallel to the
imaginary axis, a necessary condition
for stability is that the determinant
remains negative.
For Pauli-Villars type regularization, the function
$B_2(p;m^2)$ is decreasing
so that this condition is implied by the condition at
zero momentum
$$\det{\bf M}<0\ \Leftarrow\
2V''(\rho)B_2(0;m^2)+1>0\,\eqno[9].$$
 For $m$ small
$$B_2(0;m^2)=\half(d-2)K(d)m^{d-4}-a(d)\Lambda^{d-4}
+O\left(m^{d-2}\Lambda^{-2},m^2\Lambda^{d-6}\right)
.\eqno[10]$$
 $V''(\rho)$ can be expanded now around the critical $\rho$.
From the saddle point condition:
$$\rho-\rho_c =-K(d)
m^{d-2}+a(d)m^2\Lambda^{d-4}
+O\left(m^d\Lambda^{-2}\right)
+O\left(m^4\Lambda^{d-6}\right). \eqno[11]$$
$$\rho_c={1\over (2\pi)^d }\int^\Lambda {{\rm d}^d k
\over k^2}
~,~K(d)=-{\Gamma(1-d/2)
\over(4\pi)^{d/2}}~,~
a(d)= {1\over (2\pi)^d}\int{{\rm d}^d k\over k^4}
\left(1-{1\over
D^2(k^2)} \right) .
$$
The constant $a(d)$  depends on the cut-off procedure,
and one finds in Eq. [9]
for a multicritical point:
$$(-1)^n {d-2\over (n-2)!}K^{n-1}(d)m^{n(d-2)-d}
V^{(n)}(\rho_c)+1>0\,.
\eqno[12] $$
This condition is satisfied by a normal critical point
since $V''(\rho_c)>0$.
 For $n$ even it is
always satisfied  while for $n$ odd  Eq.[12] is always satisfied
above the critical
dimension and never below. At the upper-critical
dimension $2/(n-1)=d-2$
we find a condition
on the value of $V^{(n)}(\rho_c)$.

The mass-matrix has  a zero eigenvalue which
corresponds to the appearance
of a new massless excitation other than  the $\vec\Phi$
quanta (which has a mass m).
Then
$$\det{\bf M}=0\ \Leftrightarrow\ 2V''(\rho)B_2(0;m^2)+1=
0\,. \eqno[13]$$
In the two-space, the corresponding eigenvector
has components
$(\half,V''(\rho))$.
In the small $m$ limit
$V''(\rho)$ must be small and  we are
close to a multicritical point and
$$(-1)^{n-1}{d-2\over (n-2)!}K^{n-1}(d)m^{n(d-2)-d}
V^{(n)}(\rho_c)=1\,.
\eqno[14]$$
This equation has solutions only for $n(d-2)=d$,
i.e.  ,  at the critical dimension. The compatibility
then fixes the value of
  $V^{(n)}(\rho_c)=\Omega_c$.
If we take into account the leading correction to
the small $m$ behavior
we  find  instead:
$$V^{(n)}(\rho_c)\Omega_c^{-1}-1
\sim(2n-3){a(d)\over K(d)}\left({m\over\Lambda}
\right)^{4-d}.\eqno[15]  $$
This means that when $a(d)>0$ there exists
a small region
$V^{(n)}(\rho_c)>\Omega_c$ where the vector
field is massive with a small mass $m$ and the
O(N) singlet bound-state is massless. The
value $\Omega_c$ is a fixed point value.
The analysis  can be extended  to a
situation where the scalar field has a small
but non-vanishing mass $M$ and
$m$ is still small. In particular,  the neighborhood of
the special point $V^{(n)}(\rho_c)=\Omega_c$ can be  explored.
The vanishing of the determinant in Eq.[8] implies
$1+2V''(\rho)B_2(iM;m^2)=0\,.$
Because  $M$ and $m$ are small, this equation still implies
that $\rho$ is close to a point $\rho_c$ where
$V''(\rho)$ vanishes.
Since reality imposes $M<2m$, it is easy to verify
that this equation
has also solutions for only the critical dimension.

Of particular interest is the
$\eta_0 (\vec\Phi^2)^3$  theory in three dimensions,
discussed  in the past in Ref.\cite{Bardeen}.
Taking the limit $N\to \infty , \Lambda \to
\infty$ in a correlated manner with $\eta_0 \to \Omega_c$
one encounters  a manifestation of
dimensional transmutation at a nontrivial ultraviolet
fixed point.
In the massive phase described
above, scale invariance is broken only spontaneously.
Indeed, one finds that
the trace of the energy momentum tensor stays zero
 at $\eta_0=\Omega_c$:
$$<P' | \Theta_{\mu\nu} | P> = P_\mu'P_\nu  + P_\mu'P_\nu
- g_{\mu\nu}P'\cdot P + g_{\mu\nu}m^2 +
(q_\mu q_\nu -q^2g_{\mu\nu})
({m^2\over q^2} + {1\over 4})\eqno[16] $$
where $q=P'-P$.
A massless dilaton - the Goldston boson associated
with this spontaneous breaking -  appears in
the ground state spectrum
as a reflection of the Goldston realization of scale symmetry.
The normal ordering of $\Phivec^6$ induces
a  $\Phivec^4$ interaction
which guarantees the appearance
of the dilaton
pole in the physical amplitudes
as $\eta_0 \to \Omega_c$ .

The double scaling limit  results here in a theory
with an ultraviolet fixed
point  and the
following properties are found: {\bf (a)} There is a dynamical mass
$m\neq 0$ for the scalar $\Phi$ particles.
{\bf (b)} The   $\Phi - \Phi$
bound state  pseudo-Goldston boson (dilaton)
 has a finite  small mass and
the interaction term of these scalars is calculable.
This is an interesting mechanism to  produce a
low mass scalar in a
spectrum with possible phenomenological implications.
{\bf (c)} The trace of the energy momentum
tensor ${\Theta_\mu}^\mu$
is finite.  {\bf (d)} An induced $\Phi^4$ coupling appears in the
theory.

\section{DISCUSSION}
This is a short summary of  the phase structure of
O(N) symmetric quantum field theories in a singular
limit, the double scaling limit.
The main point emphasized here is that this
formal singular limit, recently discussed mainly in $d=0$
O(N) matrix models, has an intriguing  physical meaning
in $d\geq 2$ O(N) vector theories.  In this limit all orders in
${1\over N}$ are of equal importance since
at each order infrared
divergences compensate  for the decrease in
powers of ${1\over N}$. The infrared divergences are
due to the tuning of the strength of the force
$ (g \to g_C)$ between the
O(N) quanta so that a massless O(N) singlet appears in
the spectrum. At critical dimension
an interesting phase structure  is revealed,
the massless excitation
has the expected physical meaning: it is the Goldston boson
of spontaneous breaking of scale invariance - the dilaton.

\end{document}